\documentclass[10pt,conference,anonymous]{IEEEtran}

\AtBeginDocument{%
    \providecommand\BibTeX{{%
        \normalfont B\kern-0.5em{\scshape i\kern-0.25em b}\kern-0.8em\TeX}}}

\usepackage{
    latexsym,
    color,
    xcolor,
    multirow,
    array,
    listings,
    enumitem,
    multicol,
    balance,
    graphicx,
    hyperref,
    url,
    algorithmic,
    textcomp
}
\usepackage{amsmath,amsfonts}
\usepackage{fancyvrb}

\usepackage[linesnumbered,ruled,vlined]{algorithm2e}

\newcolumntype{P}[1]{>{\centering\arraybackslash}p{#1}}
\newtheorem{auxdefn}{Definition}[section]

\newtheorem{auxexample}{Example}[section]

\begin{document}

\title{Directed Test Program Generation for JIT Compiler Bug Localization}

\author{
   \IEEEauthorblockN{HeuiChan Lim}
   \IEEEauthorblockA{
       \textit{Department of Computer Science}\\
       \textit{The University of Arizona}\\
       Tucson, USA \\
       hlim1@arizona.edu}
   \and
   \IEEEauthorblockN{Saumya Debray}
   \IEEEauthorblockA{
       \textit{Department of Computer Science}\\
       \textit{The University of Arizona}\\
       Tucson, USA \\
       debray@cs.arizona.edu}
}


\maketitle

\begin{abstract}
  Bug localization techniques for Just-in-Time (JIT) compilers are based
  on analyzing the execution behaviors of the target 
  JIT compiler on a set of test programs generated for this purpose;
  characteristics of these test inputs can significantly impact the
  accuracy of bug localization.
  However, current approaches for automatic test program generation
do not work well for bug localization in JIT compilers.
This paper proposes a novel technique for automatic test program generation
for JIT compiler bug localization that is based on two key insights:
(1) the generated test programs should contain both
passing inputs (which do not trigger the bug) and failing inputs
(which trigger the bug); and (2) the passing inputs should be as similar
as possible to the initial seed input, while the failing programs should be
as different as possible from it.
We use a structural analysis of the seed program to determine
which parts of the code should be mutated for each of the passing and failing cases.
Experiments using a prototype implementation 
indicate that test inputs generated using our approach 
result in significantly improved bug localization results than existing approaches. 
\end{abstract}

\IEEEpeerreviewmaketitle

\newcommand{\smallfont}{\fontsize{8pt}{10pt}\selectfont}
\newcolumntype{P}[1]{>{\centering\arraybackslash}p{#1}}

\newcommand{\Entities}[1]{\ensuremath{{\cal E}_{P}(#1)}}

\section{Introduction}\label{sec:introduction}

Just-in-Time (JIT) compilers are widely used in modern software 
to improve the performance of interpreted systems. Bugs in JIT 
compilers can result in the generation of incorrect optimized code, 
which can then lead to exploitable security vulnerabilities
\cite{VanAmerongen2020,glazunov2021}.
The size and complexity of modern JIT compilers, and the nontrivial
manual effort needed to locate and fix such bugs,
makes it important to develop automated techniques for rapid
bug localization in JIT compilers.

Automated bug localization techniques typically rely on
analyzing the execution behavior of the JIT compiler on a set of test
inputs generated for the purpose.  The idea is to examine the set of
``program entities'' involved in the execution of the various test
inputs (e.g., functions, or source files, or data structures
manipulated) to zero in on a set of suspicious program entities
that are potential culprits for the bug.
The accuracy of this process depends in part on the size of the set of suspicious
entities identified (smaller is better), which necessarily depends on the characteristics
of the test inputs used.  This suggests that careful attention to
the set of test inputs used can help improve the accuracy
of bug localization.  This paper proposes a novel approach to generating 
test inputs that focuses on aspects of test generation that affect the
set of suspicious entities, and thus can achieve better bug localization accuracy
than existing approaches.  Our approach is based on two insights: 
(1) the generated test programs should contain both passing inputs 
(that do not trigger the bug) and failing inputs (that trigger the bug); and 
(2) the passing inputs should be similar to the original seed input, 
while the failing inputs should be as different as possible from the seed. 

Existing approaches to test input generation for JIT compiler bug localization
do not use the considerations discussed above to guide the input generation
process.  Lim and Debray's work on JIT compiler bug localization \cite{LimD21}
uses random mutation of the initial seed program, which is less effective
in minimizing the set of suspicious entities than that described here.
A body of work on bug localization in conventional compilers 
\cite{HSZHZ19,ChenMZ20,DBLP:journals/tr/ZhouJRCQ22} focuses on constructing
multiple passing test inputs but uses only a single failing input
(the original seed program).  Other researchers use the same approach to 
generate both failing and passing programs~\cite{DBLP:conf/issta/RobetalerFZO12}
and thus do not utilize the fact that treating 
passing and failing inputs differently during construction
can improve the quality of bug localization.

In order to assess the effectiveness of our approach, we developed 
a prototype tool called {\it DPGen4JIT} (Directed Program Generator 
for JIT Compiler). 
Our approach follows a series of steps to generate and select test programs. First, we generate an initial set of test programs by mutating the seed program in a non-directed manner, without specifying which nodes to mutate. Next, we analyze the initial set of programs to identify the nodes that should be mutated or avoided. We mutate the seed programs using this information to generate new test programs. Finally, we select the test programs by analyzing their similarities to the seed program. These steps enable our approach to generate and select effective test programs for bug localization.
We used the test programs created by {\it DPGen4JIT} 
to perform bug localization on 72 bugs in two widely used JIT compilers, 
Google TurboFan~\cite{turbofan} and Mozilla IonMonkey~\cite{ionmonkey}.
The results indicate that test programs generated using our approach 
lead to significantly smaller sets of suspicious entities and result in
significantly higher accuracy in bug localization than 
the existing approaches.

In summary, this paper makes the following contributions:
\begin{enumerate}
    \item it describes a novel approach to generating 
      test inputs for bug localization in JIT compilers, with the aim of reducing
      the set of suspicious program entities and improving the accuracy of
      bug localization; and
    \item it demonstrates the efficacy of our ideas using a prototype
      implementation evaluated on 72 bugs for 
    two widely used JIT compilers. The results indicate that 
    our approach leads to significantly improved accuracy for JIT compiler 
    bug localization compared to existing approaches.
\end{enumerate}

\section{Motivation}\label{sec:motivation}

\begin{figure}[h]
  \centering
  \includegraphics[width=0.9\linewidth]{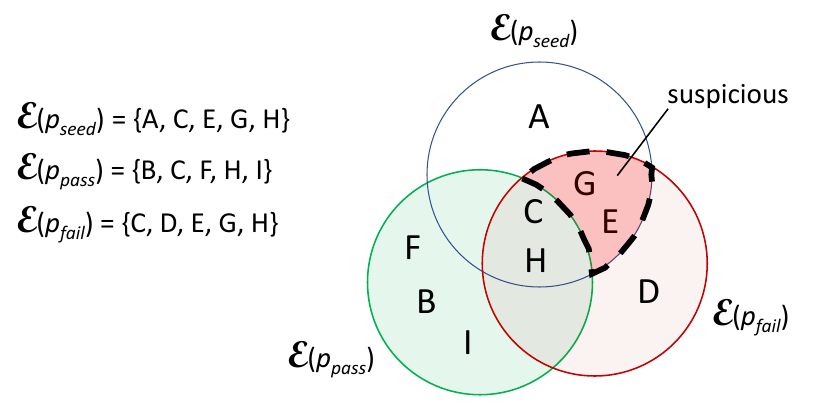}
  \caption{Buggy entity isolation example.}
  \label{fig:venn}
\end{figure}

When a buggy program $P$ under consideration is executed on a test
input $a$, the resulting execution gives rise to a set of program entities
\Entities{a} that are involved with that execution and are potentially relevant to
the buggy behavior.
Figure~\ref{fig:venn} shows a simplified example with three inputs:
the initial seed program $p_{\it seed}$
that triggers the buggy behavior, which gives rise to the
set of entities $\Entities{p_{\it seed}} = \{A, C, E, G, H\}$;
a passing input $p_{\it pass}$ (which does not
trigger the bug), with $\Entities{p_{\it pass}} = \{B, C, F, H, I\}$; and a
failing test input $p_{\it fail}$ (which triggers the bug), with
$\Entities{p_{\it fail}} = \{D, E, G, J, K\}$.  Suppose that the actual buggy entity is $G$. 
From a bug localization perspective, the set of suspicious 
entities are those that are common to all the buggy executions but not
the non-buggy ones.  In Figure \ref{fig:venn}, this is given by
$${\sf suspicious} = (\Entities{p_{\it seed}} \cap \Entities{p_{\it fail}}) - \Entities{p_{\it pass}}
  = \{G, E\}.$$
The ground truth buggy entity $G$ is in this set, but it also contains
a (spurious) non-buggy entitiy $E$, which can potentially impact the
accuracy of the bug localization process.  Ideally, therefore, the
test inputs should be such that the resulting set of suspicious
entities is as small as possible.  

More generally, suppose that we have a set of passing test inputs
${\bf P}_{\it pass}$ that do not trigger the bug
and a set of failing inputs ${\bf P}_{\it fail}$ that trigger the bug.
Reasoning as above, the set of suspicious entities is given by
$${\sf suspicious} =
  (\bigcap_{a \in {\bf P}_{\it fail}} \Entities{a}) - (\bigcup_{b \in {\bf P}_{\it pass}} \Entities{b}).$$
To minimize the size of this set, we should minimize the size of
the intersection and maximize the size of the union.
In other words, we should $(i)$ make the entities arising from
inputs in ${\bf P}_{\it fail}$ as dissimilar as possible;
and $(ii)$ make the entities arising from inputs in 
${\bf P}_{\it pass}$ as similar as possible to the initial failing seed input.
This is the basic idea guiding the approach to test program generation
described here.
 
One issue that arises here is that the set of program entities
arising from executing a program on a given input become known only once it
has been executed, while we have to construct the test inputs prior to execution.
To address this, we use structural similarity between the
input test programs as a proxy for the similarity or difference between
the entities arising from their execution.

\section{Background: JIT Compilers and Test Oracles}\label{sec:background}

JIT compilers are used to improve the performance of interpreted systems
by dynamically optimizing the interpreted byte-code to more efficient native code.
The JIT compiler is tightly coupled with an interpreter that converts the
input program to a machine-independent byte-code representation,
which is then interpreted.
The runtime system monitors the execution of the byte-code as it is interpreted,
and frequently executed code fragments are passed to the JIT compiler to be
compiled to native code.
The generated native code contains checks to ensure that assumptions made
during optimization, e.g., about the types of variables, are not violated.
If an assumption is found to not hold, the corresponding code fragment is
``de-optimized'' back to the original byte-code.

This system structure provides a convenient test oracle for JIT compiler bugs.  
We define the execution of a JIT compiler to be buggy if the observable
behavior of the input program is different when it is interpreted than when
it is JIT-compiled.  For any given input $p$, therefore, we can determine whether
$p$ triggers a JIT compiler bug by comparing its observable execution behavior with
JIT-compilation enabled with that with JIT-compilation disabled.

\section{System Design}\label{sec:research}

\begin{figure*}
  \centering
  \includegraphics[width=0.8\linewidth]{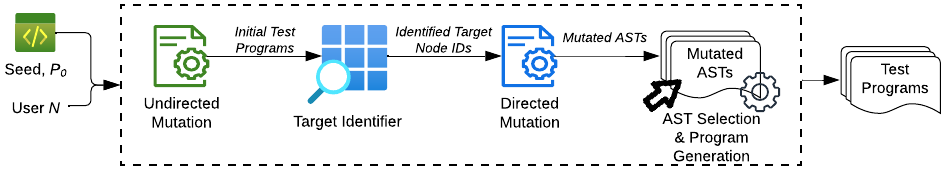}
  \caption{The architecture of {\it DPGen4JIT}.}
  \label{fig:pipeline}
\end{figure*}


\subsection{Terminology}\label{sec:terminlologies}
We use the following terminology in the paper:
\begin{itemize}
    \item {\bf Test Program} is a piece of code, such as JavaScript, 
    that serves as input to another software, like a JIT compiler, 
    to analyze and observe the behavior of the software.
    \item {\bf Seed (Test) Program} is an original test program known 
    to trigger a bug in the JIT compiler.
    \item {\bf Failing (Test) Program} is a test program that replicates 
    the buggy behavior of the seed test program when executing with 
    the same target software being tested.
    \item {\bf Passing (Test) Program} is a test program that no longer 
    triggers a bug in the target software being tested.
    \item {\bf Valid Program} is a test program that does not fail due 
    to a non-bug-related issue (e.g., a syntax error). For example, valid programs do not fail 
    at the parser because of the mutation we applied to the seed program.
\end{itemize}

\subsection{Overview}\label{sec:overview}

We need a seed test program known to trigger a bug in the JIT compiler to generate passing and failing test programs, as these programs are mutations of the seed. We can obtain the seed test programs from the bug reports. When developers find a new bug, they submit a report with the ``proof-of-concept'' (PoC) code.\footnote{
  Vendors require bug reports to specify how the bug can be
  reproduced (e.g., see \cite{bugzilla-guidelines,chromium-security-bugs}).
  For JIT compiler bugs, this translates to providing a PoC input that triggers
  the buggy behavior.}
We also assume that the user specifies the number of test inputs to generate.

Given a seed test program $P_0$ and the number $N$ of test programs to generate, we take the following steps illustrated in Figure~\ref{fig:pipeline} to generate and select the passing and failing programs for the bug localization:
\begin{enumerate}
	\item {\bf Undirected Test Program Generation} (Section \ref{sec:initial}).
          Generate test programs by traversing $P_0$'s abstract syntax tree (AST)
          and attempting to mutate each node.  Each node mutation results in a new
          program.  Some of the resulting mutated programs will continue to trigger 
        the JIT compiler bug while others will not.
	\item {\bf Target Identification} (Section \ref{sec:target-identification}). 
          Use the programs generated in Step 1 to identify:
          $(i)$ which programs no longer trigger the JIT compiler bug; and $(ii)$
          for each such non-bug-triggering program $P$, which AST node in $P_0$ was mutated
          to obtain $P$.
	\item {\bf Directed Mutation} (Section \ref{sec:target-mutation}).
          Using the information from Step 2 about which AST nodes
          affect triggering the JIT compiler bug, 
          the tool targets the nodes to mutate in order to generate test programs
          that either trigger (failing inputs) or do not trigger the bug
          (passing inputs).
	\item {\bf Test Program Selection} (Section \ref{sec:selection}).
        Given the passing and failing inputs generated in Step 3,
        the tool selects the input programs used for bug localization.
\end{enumerate}

\subsection{Mutation Policy}\label{sec:policy}

To produce a valid test program, we use the following rules when 
mutating the abstract syntax tree (AST) of a program:

\begin{enumerate}
    \item the mutated program must be syntactically correct; and
    \item the mutation must not violate the semantics of the language, 
    and the mutated program should be semantically similar to the seed 
    program.
\end{enumerate}

We use language specifications to ensure the validity 
of mutations made to the program. E.g.,
for JavaScript we consulted the ECMA~\cite{ecmastandard} 
and Mozilla~\cite{mozillaJSRef} language specifications.

The purpose of adhering to these guidelines is not only to produce 
a valid program but also to generate test programs that traverse 
a comparable execution path to the original seed program. 
The rationale behind this is that by examining the similarities 
and differences between the newly generated test programs and 
the seed program, explained in Section~\ref{sec:motivation}, 
which share comparable syntax and meaning, 
bug localization methods can accurately identify and isolate 
the buggy components from the failing execution, 
i.e., the seed program's execution.

\begin{table}[h]
\caption{Categorized rules}\label{tbl:rules}
\begin{center}
\begin{tabular}{|l|p{6cm}|}\hline
  {\bf Rules} & {\bf Description}  \\ \hline\hline
  Operators & Operators are replaced with another operator within 
  the same group. For example, binary operators are replaced 
  with another binary operator, unary operators with another 
  unary operator, etc. \\ \hline
  Built-in Methods & Built-in methods are substituted with 
  another built-in method from the same group with 
  the same parameters.\\ \hline
  Values & Values are replaced according to their types. 
  For instance, integer values are substituted with another 
  integer value, float values with another float value, 
  strings with another string, and so on. \\ \hline
\end{tabular}
\end{center}
\end{table}

\subsubsection{Syntax Rule}\label{sec:syntax}


We only make changes to the syntax tree through {\it replacement} 
to generate test programs that do not violate the syntax rules of
the language.
In particular, we do not add or remove nodes from the tree.

The rules in our mutation policy are summarized in Table~\ref{tbl:rules}, 
where each category is further sub-categorized. 
For instance, we have a separate category for built-in functions 
for string operations and another for integer operations. 
Additionally, different syntaxes must segment built-in functions 
to differentiate between operations with the same number of parameters. 
For example, the square root function receives one argument, 
while the power function receives two arguments in 
the JavaScript language. 

\subsubsection{Semantic Rule}\label{sec:semantic}

To ensure that the newly generated programs maintain semantic 
similarity with the seed program, we ensure that any modifications
to the AST preserve the node's type. This implies that we only 
substitute a literal with another literal, a binary arithmetic 
operator with another binary arithmetic operator, an integer 
constant with another integer constant, and so on.
In addition to the rules outlined in 
Table~\ref{tbl:rules}, we further categorized the operators
and functions based on the type of value they operate on and 
return. For instance, we grouped unary operators like {\tt $+$}, 
{\tt $-$}, and {\tt $\sim$}, which return a number, separately 
from {\tt $!$}, which returns a logical value.
\begin{figure*}[h]
  \centering
  \begin{tabular}{l|l|l|l} 
    1 {\bf function} foo(x) \{ & {\bf function} foo(x) \{ & {\bf function} foo(x) \{ & {\bf function} foo(x) \{\\
    2 \hspace{0.3cm}let y = x + -0; & \hspace{0.3cm}let y = x + \fbox{+}0; & \hspace{0.3cm}let y = x + -0; & \hspace{0.3cm}let y = x \fbox{*} -0;\\
    3 \hspace{0.3cm}let z = Math.max(x,y); & \hspace{0.3cm}let z = Math.max(x,y); & \hspace{0.3cm}let z = Math.\fbox{min}(x,y); & \hspace{0.3cm}let z = Math.max(x,y);\\
    4 \hspace{0.3cm}{\bf if} (y == x \&\& y == z) & \hspace{0.3cm}{\bf if} (y == x \&\& y == z) & \hspace{0.3cm}{\bf if} (y == x \&\& y == z) & \hspace{0.3cm}{\bf if} (y \fbox{!=} x \&\& y == z) \\
    5 \hspace{0.6cm}{\bf return} true; & \hspace{0.6cm}{\bf return} true; & \hspace{0.6cm}{\bf return} true; & \hspace{0.6cm}{\bf return} \fbox{false};\\
    6 \hspace{0.3cm}{\bf return} false; & \hspace{0.3cm}{\bf return} false; & \hspace{0.3cm}{\bf return} false; & \hspace{0.3cm}{\bf return} false;\\
    \multicolumn{1}{c}{${\bf Seed}$} & \multicolumn{1}{c}{${\bf P_1}$} & \multicolumn{1}{c}{${\bf P_2}$} & \multicolumn{1}{c}{${\bf P_3}$}\\
  \end{tabular}
  \caption{Examples of generated programs from mutating the seed program.}
  \label{fig:examples}
\end{figure*}

Figure~\ref{fig:examples} shows examples of new programs 
generated from mutating the seed program. The unary operator 
{\tt $-$} is replaced with the unary operator {\tt $+$} at
line 2 for program $P_1$. The built-in function {\tt $max$} 
is replaced with the built-in function {\tt $min$} at line 3 
for program $P_2$. For program $P_3$, multiple locations are 
mutated. Arithmetic operator {\tt +} is replaced with 
{\tt *} operator (line 2), {\tt ==} operator is replaced 
with {\tt !=} (line 4), and the boolean value {\it true}
is replaced with the value {\it false} (line 5).

\subsection{Undirected Test Program Generation}\label{sec:initial}

Given a seed test program, we generate a new set of programs by mutating the seed without any directions on which nodes to target. This step aims to create various new programs different from the seed; some continue to trigger a bug in the JIT compiler, while others do not. This step can proceed in two options: (1) generate variants using the existing random test program generators (e.g., Lim and Debray's fuzzer~\cite{LimD21}, etc.); or (2) generate variants using our undirected test program generator. However, our undirected test program generator has an advantage over existing random test program generators. This is because it systematically traverses all nodes in the AST and attempts to mutate them, resulting in a wider range of generated programs. In contrast, random AST mutation relies on the random selection of AST nodes, which can limit the diversity of the generated programs.

Our undirected test program generator strictly follows the rules described in Section~\ref{sec:policy}.
Let ${\it mutate}(A, i)$ denote the mutated AST $A$ at $i^{th}$ node. Given the seed program $P_0$, the tool first constructs the AST ($AST_0$). Algorithm~\ref{algo:random} shows the steps of generating random test programs from the seed AST. First, the ${\it undirected}$ function prepares an empty {\it UNDIRECTED} set to hold the newly generated test program ASTs (line 2). The tool iterates as the same number of nodes in the seed AST aiming to mutate all the nodes to generate as variety as possible new programs that differ from each other (lines 4 - 8). Mutate the $i^{th}$ node of copied AST (lines 5 - 6), then add it to the {\it UNDIRECTED} set (line 7). Finally, a set of new test program ASTs, {\it UNDIRECTED}, is returned (line 9).
\begin{algorithm}[h]
    \DontPrintSemicolon
    \SetKwInput{KwData}{Input}
    \KwResult{{\it UNDIRECTED}: Set of initial test program ASTs.}
    \SetKwFunction{newphase}{newphase}
    \KwData{$AST_0$: AST of the seed test program.}
    \SetKwFunction{FMain}{{\it undirected}}
    \SetKwProg{Fn}{function}{:}{}
    \Fn{\FMain{$AST_0$}}{
        ${\it UNDIRECTED} \leftarrow \emptyset$\;
        $i \leftarrow 1$\;
        \While{$i < size(AST_0)+1$} {
            $ast\_copy \leftarrow copy(AST_0)$\;
            $ast\_copy \leftarrow {\it mutate}(ast\_copy, i)$\;
            ${\it UNDIRECTED} \leftarrow {\it UNDIRECTED} \cup \{ast\_copy\}$\;
            $i \leftarrow i + 1$\;
        }
        \Return {\it UNDIRECTED}\;
    }
    \caption{Undirected test program generation}\label{algo:random}
\end{algorithm}

\subsection{Target Identification}\label{sec:target-identification}

This step aims to identify the specific AST nodes corresponding to 
a section(s) of the seed test program that triggers the bug in the JIT compiler.
Given the seed test program $P_0$ and set of randomly generate test programs 
${\it UNDIRECTED} = \{U_1, U_2, ..., U_{H}\}$, where $U_i$ is a randomly generated 
test program and $H$ is the number of programs, we take the following steps 
to identify the specific nodes of the seed program AST:

\begin{enumerate}
    \item Identify passing programs from ${\it UNDIRECTED}$.
    \item Construct ASTs of seed program $P_0$ and test programs.
    \item Compute the differences between the seed program AST and
    the ASTs of the passing test programs. The differences are 
    the target nodes to mutate.
\end{enumerate}

\subsubsection{Identify Passing Programs with the Test Oracle}\label{sec:passing-programs}

To determine whether a JIT compiler's execution is bug-free, we check whether 
the behavior of the test program is consistent with or without JIT compilation. 
If the behaviors of the interpreted and JIT-compiled versions of the program 
differ, then it is considered buggy. Otherwise, it is non-buggy.

\subsubsection{Construct ASTs of Programs}\label{sec:ast-construct}

This step constructs abstract syntax trees for all passing programs 
and the seed program.

\subsubsection{Compute the Differences} 
\begin{figure}[h]
  \centering
  \includegraphics[width=\linewidth]{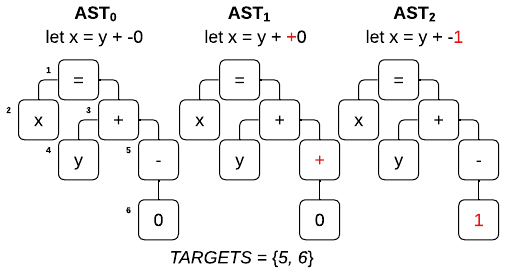}
  \caption{Finding different AST nodes.}
  \label{fig:ast-difference}
\end{figure}

To identify differences between the syntax tree of a seed program (${\it AST}_0$) 
and those of the passing programs, we utilize a process that
involves computing the AST difference. Figure~\ref{fig:ast-difference}
demonstrates this process, which involves identifying different AST nodes. 
For instance, if we consider the seed program AST ${\it AST}_0$ from the JavaScript
program $let\ x = y + -0$, and compare it to two other ASTs (${\it AST}_1$ and ${\it AST}_2$)
from mutated programs, we identify the AST nodes of ${\it AST}_0$ that differ 
from ${\it AST}_1$ and ${\it AST}_2$. In this example, the $5^{th}$ and $6^{th}$ nodes 
are different, so we add AST node IDs $5$ and $6$ to the ${\it TARGETS}$ set. 
To perform this comparison, we use the Needleman-Wunsch sequence alignment 
algorithm~\cite{DBLP:conf/dasc/NaiduN16} in our implementation.

\subsection{Directed Mutation}\label{sec:target-mutation}

Given the set of seed program's AST node ids, which we know are related to 
a bug in a JIT compiler, this step aims to generate two sets of passing and
failing programs. Generating passing programs is based on the idea that the
difference between two very similar executions, one is buggy and another 
is not, is the part that holds information on the bug as discussed 
in Section~\ref{sec:motivation}. Thus, we generate passing programs 
highly similar to the seed test program by making a minor 
mutation to the copy of the seed AST specifically targeting the identified
AST nodes in the earlier step. In contrast, generating failing programs
is based on the intuition that analyzing the commonality between the two very 
different execution while both are buggy allows us to identify the bug in 
the JIT compiler. To generate the failing programs, we mutate the nodes of
the seed AST copy except the identified target nodes.


\subsubsection{Generating Passing Programs}\label{sec:gen-passings}
Algorithm~\ref{algo:generate-passings} generates a set of passing program ASTs 
by mutating the seed test program AST, ${\it AST}_0$. The function takes the number
of programs to generate, $N$, and the set of AST node IDs to mutate, ${\it TARGETS}$, 
as inputs.
The set ${\it PASSINGS}$, initialized to $\emptyset$, holds the generated ASTs,
the variable ${\it cur\_ast}$ keeps track of the number of such ASTs, and
${\it next\_target}$ refers to the next target node to mutate.

While iterating the loop $N$ times, the function makes a copy of 
the original AST and mutates the target node identified by the ${\it TARGETS}$ 
set using the ${\it mutate}$ function. If the mutated AST is not 
in the ${\it PASSINGS}$ set, it is added to the set and the next target node 
is selected for mutation. If the mutated AST is already present in 
the ${\it PASSINGS}$ set, the function proceeds to the next target node 
without adding the AST to the set. If all target nodes have been 
mutated, the function starts again from the first target node 
in the ${\it TARGETS}$ set. The function returns the ${\it PASSINGS}$ set 
when the desired number of ASTs have been generated.

\begin{algorithm}[h]
    \DontPrintSemicolon
    \SetKwInput{KwData}{Input}
    \SetKwFunction{newphase}{newphase}
    \KwData{${\it AST}_0$: AST of the seed test program.}
    \KwData{${\it N}$: User-specified number.}
    \KwData{${\it TARGETS}$: Set of target AST node IDs.}
    \KwResult{${\it PASSINGS}$: Set of new ASTs.}
    \SetKwFunction{FMain}{{\it generate\_passings}}
    \SetKwProg{Fn}{function}{:}{}
    \Fn{\FMain{$N$, ${\it AST}_0$, ${\it TARGETS}$}}{
        ${\it PASSINGS} \leftarrow \emptyset$\;
        ${\it cur\_ast} \leftarrow 1$\;
        ${\it next\_target} \leftarrow 1$\;
        \While {${\it cur\_ast} \leq N$} {
            ${\it ast\_copy} \leftarrow copy({\it AST}_0)$\;
            ${\it target\_id} \leftarrow {\it get\_target}({\it TARGETS}, {\it next\_target})$\;
            ${\it ast\_copy} \leftarrow {\it mutate}({\it ast\_copy}, {\it target\_id})$\;
            \If{${\it ast\_copy} \notin {\it PASSINGS}$} {
                add {\it ast\_copy} to {\it PASSINGS}\;
                ${\it next\_target} \leftarrow {\it next\_target} + 1$\;
            }
            \If{${\it next\_target} > {\it size}({\it TARGETS})$} {
                ${\it next\_target} \leftarrow 1$\;
            }
            ${\it cur\_ast} \leftarrow {\it cur\_ast} + 1$\;
        }
        \Return ${\it PASSINGS}$\;
    }
    \caption{Generating passing program ASTs.}\label{algo:generate-passings}
\end{algorithm}

\subsubsection{Generating Failing Programs}\label{sec:gen-fails}
Algorithm~\ref{algo:generate-fails} generates a set of 
failing program ASTs by mutating ${\it AST}_0$. 
It takes the same inputs as the ${\it generate\_passings}$ function. 
The set ${\it FAILS}$, initialized to $\emptyset$, holds the generated ASTs,
while the variable ${\it cur\_ast}$ keeps track of the number of such ASTs.

For each iteration, the function makes a copy of the original AST 
and mutates the nodes that are not related to the bug in the JIT compiler,
i.e., the node id is not in the ${\it TARGETS}$ set, using the ${\it mutate}$ function.
Unlike the ${\it generate\_passings}$ function, the ${\it generate\_fails}$ function 
does not break out of the loop after mutating the target nodes. 
It fully traverses the AST to mutate all relevant nodes 
to generate new failing programs that are different from the seed program. 
If the mutated AST is not in the ${\it FAILS}$ set, it is added to the set. 
If the mutated AST is already present in the ${\it FAILS}$ set, 
the function proceeds to the next mutation without adding 
the AST to the set. The function returns the ${\it FAILS}$ set.

\begin{algorithm}[h]
    \DontPrintSemicolon
    \SetKwInput{KwData}{Input}
    \SetKwFunction{newphase}{newphase}
    \KwData{${\it AST}_0$: AST of the seed test program.}
    \KwData{${\it N}$: User-specified number.}
    \KwData{${\it TARGETS}$: Set of target AST node IDs.}
    \KwResult{${\it PASSINGS}$: Set of new ASTs.}
    \SetKwFunction{FMain}{{\it generate\_fails}}
    \SetKwProg{Fn}{function}{:}{}
    \Fn{\FMain{$N$, ${\it AST}_0$, ${\it TARGETS}$}}{
        ${\it FAILS} \leftarrow \emptyset$\;
        ${\it cur\_ast} \leftarrow 1$\;
        \While {${\it cur\_ast} \leq N$} {
            ${\it ast\_copy} \leftarrow {\it copy}({\it AST}_0)$\;
            \For {${\it node} \in {\it ast\_copy}$} {
                \If{${\it ID}({\it node}) \notin {\it TARGETS}$} {
                    ${\it ast\_copy} \leftarrow {\it mutate}({\it ast\_copy}, {\it ID}({\it node}))$\;
                }
            }
            \If{${\it ast\_copy} \notin {\it FAILS}$} {
                add {\it ast\_copy} to {\it FAILS}\;
            }
            ${\it cur\_ast} \leftarrow {\it cur\_ast} + 1$\;
        }
        \Return ${\it FAILS}$\;
    }
    \caption{Generating failing program ASTs.}\label{algo:generate-fails}
\end{algorithm}

\subsection{Test Program Selection}\label{sec:selection}

The final step of {\it DPGen4JIT} is to convert $N$ number of ASTs, 
which are selected from the ${\it PASSINGS}$ and ${\it FAILS}$ sets, to
source code (e.g., JavaScript code) that can be used by the bug localizer
as input to the JIT compiler.  
The AST selection guarantees our idea further that the failing test 
programs to use in the bug localization analysis should be as different 
as possible from the seed. In contrast, the passing test programs should 
be similar as possible. We use {\it Jaccard similarity} to compute
the similarity between the seed program AST and mutated ASTs.
Let ${\it Nodes}(T)$ denote a set of nodes in abstract syntax tree $T$.
\begin{align*}
  Sim(T_0, T_i) = \frac{\mid {\it Nodes}(T_0) \cap {\it Nodes}(T_i) \mid}{\mid {\it Nodes}(T_0) \cup {\it Nodes}(T_i) \mid}
\end{align*}

$Sim(T_0, T_i)$ denotes the similarity value between the set of nodes
in $T_0$ and $T_i$, where $T_0$ denotes the seed test program AST and $T_i$ 
denotes the AST of a new test program to compare the similarity with
the seed.

Given a (user-specified) value $N$ of the number of test programs to generate,
our current implementation constructs $N/2$ passing and $N/2$ failing programs
for bug localization.
To construct the set of passing test inputs,
we calculate the similarity between each AST in the ${\it PASSINGS}$ set, 
and the seed AST, $T_0$, sort the results in descending order and 
select the top $N/2$ ASTs; these are the passing programs most similar to
the initial seed input.  To construct the set of failing inputs,
we calculate the similarity between each AST in the ${\it FAILS}$ set, 
and the seed AST, $T_0$, sort the results in ascending order and 
select the top $N/2$ ASTs; these are the failing programs most dissimilar
to the initial seed input.
The resulting ASTs are then written out as source programs that can
be provided as input to the JIT compiler.

\subsection{Summary}\label{sec:summary}
\begin{algorithm}[h]
    \DontPrintSemicolon
    \SetKwInput{KwData}{Input}
    \SetKwFunction{newphase}{newphase}
    \KwData{$P_0$: The seed test program.}
    \KwData{${\it N}$: User-specified number.}
    \SetKwFunction{FMain}{{\it generate\_fails}}
    \SetKwProg{Fn}{function}{:}{}
    \Begin {
        ${\it AST}_0 \leftarrow {\it get\_ast}(P_0)$\;
        ${\it UNDIRECTED} \leftarrow {\it undirected}({\it AST}\_0)$\;
        ${\it TARGETS} \leftarrow {\it target\_identification}({\it UNDIRECTED}, P_0)$\;
        ${\it PASSINGS} \leftarrow {\it generate\_passings}(N, AST_0, {\it TARGETS})$\;
        ${\it FAILS} \leftarrow {\it generate\_fails}(N, AST_0, TARGETS)$\;
        ${\it SELECTED} \leftarrow {\it select}({\it PASSINGS}, {\it FAILS}, N)$\;
        ${\it PROGRAMS} \leftarrow {\it convert}(SELECTED) \cup \{P_0\}$\;
        \Return ${\it PROGRAMS}$
    }
    \caption{Overall algorithm}\label{algo:overall}
\end{algorithm}

Our approach is summarized in Algorithm~\ref{algo:overall}, which provides an outline of the entire process. We begin with a seed test program $P_0$ and the user-specified value $N$, which determines the number of test programs to be generated. Initially, we create the abstract syntax tree of $P_0$. Next, we randomly mutate the AST of $P_0$ to generate test programs in an undirected manner. We then examine these test programs to determine which AST nodes to mutate in order to produce passing or failing test programs. Using this information, we generate ASTs for both passing and failing test programs. We then select the most appropriate ASTs: failing test programs that are substantially different from the seed, and passing test programs that are similar to it. Finally, we convert these selected ASTs into actual programs and return them.

\section{Evaluation}\label{sec:evaluation}

We conducted experiments using our proposed approach, 
which we implemented in a prototype tool called {\it DPGen4JIT}.
The experiments were performed on a machine with 32 cores (@ 3.30 GHz) 
and 1TB of RAM, running Ubuntu 20.04.1 LTS. We used the esprima-python
library~\cite{esprima} to generate ASTs for JavaScript code
and escodegen~\cite{escodegen} to converting mutated ASTs to JavaScript code.
Source code and data for {\it DPGen4JIT} can be found at 
\url{https://github.com/hlim1/DPGen4JIT}.

\subsection{Research Questions}

Our experimental evaluation considered the following research questions:
\begin{enumerate}
    \item How effective in {\it DPGen4JIT} in
      reducing the number of suspicious entities considered for
      bug localization?
      
    \item
      How does {\em DPGen4JIT} compare with existing approaches for
      test input generation with regard to reducing the number of
      non-suspicious entities?
      
    \item How does the use of programs generated by {\it DPGen4JIT} 
    impact the accuracy of bug localization compared to 
    existing input generation techniques?
\end{enumerate}

\subsection{Benchmarks and Target Systems}

We evaluated the efficacy of {\it DPGen4JIT} on 72 optimization bugs from two of 
the most widely used JIT compilers: TurboFan (V8; Google) and 
IonMonkey (SpiderMonkey; Mozilla).
Out of these bugs, 21 were reported on the vendors' websites, 
while 51 were synthetic bugs that had similar characteristics to
the reported bugs and were plausible in real-world situations. 
The criteria used to select the bug reports were: 
(1) the bug had to be in the JIT compiler's optimizer;
(2) the buggy behavior had to be replicable, with the same behavior
observable in the provided PoC code and options;
(3) it had to be possible to identify an incorrectly optimized IR node 
from the JIT compiler source code, with the buggy function accessing 
the IR node to either manipulate the property or create a new node; and
(4) the bug had to be marked as ``fixed.'' The last criterion enabled us 
to use the fixed code, together with developer comments, to
obtain ground truth information about the buggy code and thereby
evaluate the accuracy of our analysis.
Due to space limitations, we omit a detailed description 
of the bugs here, but it is available in the data submitted 
with the paper.

We generated binary executables for two widely-used JavaScript engines,
namely V8 (Google) and SpiderMonkey (Mozilla), using the debugging settings
commonly employed by developers. Subsequently, we executed the input 
JavaScript code for each bug with the relevant executable options. 
For instance, we used the \texttt{--fast-warmup} option to expedite 
the warm-up phase of the SpiderMonkey JIT compiler.

\subsection{Bug Localization}\label{sec:bug-localization}

To perform bug localization, we followed the approach proposed by
Lim and Debray~\cite{LimD21}. This approach analyzes the execution
trace of a JIT compiler and extracts information on the intermediate 
representation (IR) that the JIT compiler constructs and optimizes. 
Based on this information, the approach constructs its own abstract 
model that corresponds to the concrete IR. First, we generate a set 
of programs using our prototype tool {\it DPGen4JIT} from 
a proof-of-concept (PoC) code (i.e., seed program). We construct 
abstract models for each program, including the seed program. 
Then, we use the Ochiai formula~\cite{AbreuOchiai}, which is one of 
the most well-known Spectrum-Based Fault Localization (SBFL) formulas~\cite{DBLP:journals/corr/SouzaCK16}, to calculate 
the suspicious values for each executed JIT compiler instruction 
on the IR during optimization. 
\begin{align*}
    Sus(I) = \frac{I_{ef}}{\sqrt{(I_{ef}+I_{nf})(I_{ef}+I_{ep})}}
\end{align*}

The suspicious value of an executed instruction $I$ is denoted 
as $Sus(I)$. $I_{ef}$ and $I_{nf}$ represent the number of failing 
programs that executed and did not execute the instruction $I$, respectively. 
$I_{ep}$ represents the number of passing programs that executed 
the instruction $I$. Subsequently, the executed instructions are sorted 
in descending order based on their suspicious values and then aggregated 
at a function level (i.e., the final output from the bug localization 
analysis is a file holding the ranking of suspicious functions). 
In addition, we ran the experiment (i.e., from generating new sets
of test programs to bug localization) three times for each bug.
We identified a ground truth function and found its rank position 
in each of the three rankings. Then we calculated the median of 
the three rank positions to obtain the final rank position of 
the ground truth item.
\cite{HSZHZ19,ChenMZ20,DBLP:journals/tr/ZhouJRCQ22}.


We apply the Top-{\em n} metric, where {\em n = 1, 5, 10, 20}, 
to measure the accuracy of bug localization result.
This metric counts the number of bugs where the ground truth bug
is localized to within the top {\em n} positions in the ranking determined
by the localization algorithm; smaller values of $n$ correspond to
greater accuracy.
For example, if the ground truth bug location is ranked third in the ranking
produced by the localization algorithm,, we consider the bug to be localized
within the Top-5. 
Kochhar {\em et al.} use this metric to assess developer preferences for
bug localization tools \cite{kochhar2016practitioners}: a ranking within the
Top-5 is regarded as ``accurate," while a ranking within the Top-10
is deemed ``acceptable." 

\subsection{Impact on the Number of Suspicious Functions}\label{sec:num-suspicious}
\begin{figure}
  \centering
  \includegraphics[width=0.9\linewidth]{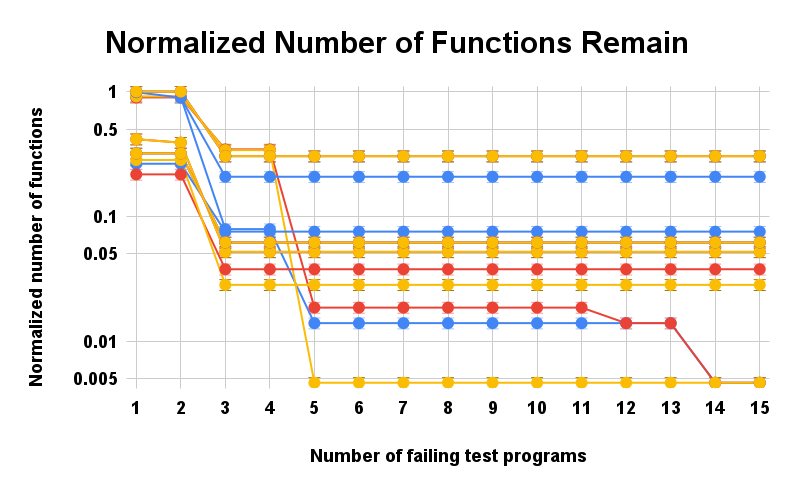}
  \caption{Normalized impact of different numbers of passing and failing input programs on the number of suspicious functions considered for bug localization (smaller is better). The line colors indicate different numbers of passing programs used, i.e., blue is 5, red is 10, and yellow is 15.}
  \label{fig:suspicious-funcs}
\end{figure}

The intuition discussed in Section~\ref{sec:motivation} suggests that
increasing the number of passing and failing inputs can reduce the
number of suspicious entities considered during bug localization.
Figure~\ref{fig:suspicious-funcs}
shows the results of an experiment to evaluate this
({\bf Research Question 1}).  We selected 5 bugs out of 72 bugs 
at random and, for each bug, computed the number of suspicious 
functions obtained using $m$ passing and $n$ failing test programs, with
$m \in \{5, 10, 15\}$ and $1 \leq n \leq 15$.
The results demonstrate the improvement 
of the number of suspicious functions as additional failing test programs 
are introduced in the analysis.
The $x$-axis shows the number of failing inputs, while the red, yellow, and blue
lines in Figure~\ref{fig:suspicious-funcs} correspond to 5, 10, and 15
passing inputs respectively.
The $y$-axis shows, on a log scale, the (normalized) number of functions that are 
not eliminated as non-suspicious for the purposes of bug localization and which
are therefore candidates for consideration in bug localization.  Since the number
of functions executed for the different bugs are very different, we normalize
the data with respect to the number of functions in the original seed input
for each bug, so as to make it easier to see what fraction of those functions
is being eliminated using the test generation algorithm of {\em DPGen4JIT}.
A normalized value closer to 1 on the y-axis 
indicates that the number of remaining functions is close to the total 
number of functions in the seed program, while a value closer to 0 
indicates a better outcome with fewer remaining functions.  A logarithmic
scale is used for the $y$-axis to produce a clearer visualization of the data
for small values of $y$, i.e., when only a relatively few suspicious functions remain.

Figure~\ref{fig:suspicious-funcs} shows that the number of suspicious functions
declines sharply as additional failing test programs are introduced, and then 
tends to plateau after the third to fifth test program. This observation 
suggests that having multiple failing test inputs in addition 
to the seed test program can be highly effective in narrowing down 
the set of suspicious functions.

\begin{figure}
  \centering
  \includegraphics[width=0.9\linewidth]{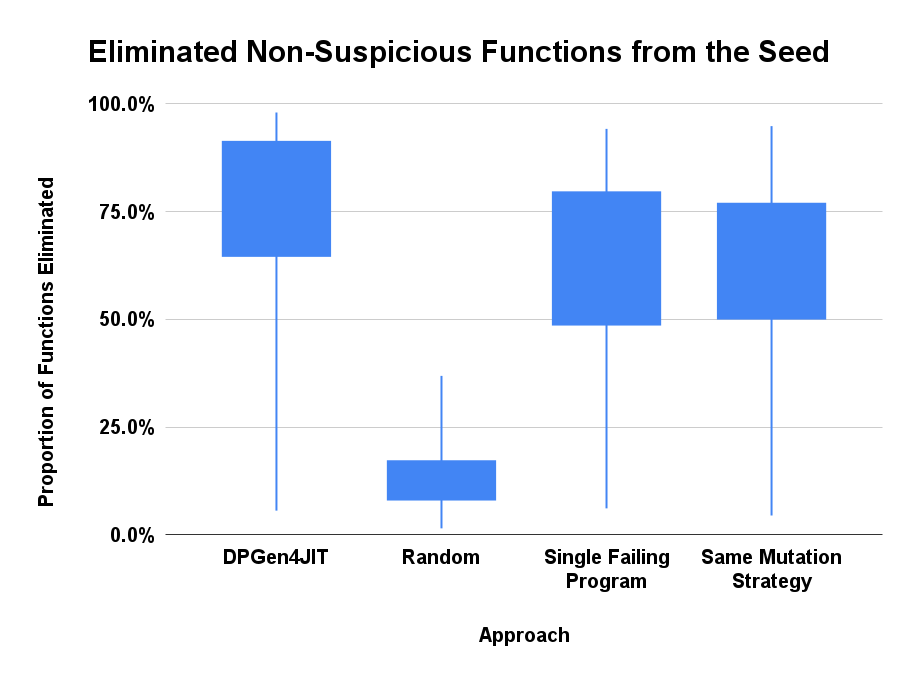}
  \caption{The effectiveness of different approaches to test input generation on proportion of non-suspicious functions eliminated}
  \label{fig:eliminated-compare}
\end{figure}


We next considered how the reduction in suspicious functions obtained
using {\em DPGen4JIT} compares
with existing approaches ({\bf Research Question 2}).
We compared {\em DPGen4JIT} with three existing approaches:
Random \cite{LimD21};
Single Failing Program, which uses a single failing input together 
with multiple passing inputs; and Same Mutation Strategy, 
which uses the same mutation strategy for both passing and failing inputs.  
The following steps were taken to achieve this. 
(1) we computed the suspicious set of functions 
using the formula outlined in Section~\ref{sec:motivation}. 
(2) we calculated the number of eliminated functions by subtracting 
the remaining functions from the total number of functions in 
the seed execution. (3) we determined the proportion of the number 
of functions eliminated to the number of seed functions. We used this 
proportion as a standardized measure to compare the effectiveness of 
different approaches across different bugs, irrespective of the total 
number of functions in each seed execution.

In Figure~\ref{fig:eliminated-compare}, the proportion of eliminated 
functions compared to the initial number of functions is shown for 
different approaches. The boxes represent the median and interquartile 
range, while the whiskers indicate the minimum and maximum percentage 
of eliminated functions. 
The results demonstrate that our approach outperforms the other 
approaches regarding the proportion of eliminated non-suspicious functions.
Our approach eliminated a median of 91.2\% of functions, 
while the random test program generation approach only removed 17.1\%, 
the single failing program approach eliminated 79.5\%, and the same 
mutation strategy approach eliminated 76.8\%.

\subsection{Impact on Bug Localization Accuracy}\label{sec:effectiveness}
\begin{table*}[h]
\caption{Suspicious Function Rankings Result}\label{tbl:ranking-result}
\begin{center}
\begin{tabular}{|c|c|c|c|c|c|c|}
    \hline
        {\bf System}
        & {\bf Total bugs}
        & {\bf Top-1} & {\bf Top-5} & {\bf Top-10} & {\bf Top-20} \\
    \hline\hline
    V8 & 37 & 4 (10\%) & 9 (24.3\%) & 15 (40.5\%) & 22 (59.5\%)\\
    \hline
    SpiderMonkey & 35 & 14 (40\%) & 22 (62.9\%) & 24 (68.6\%) & 28 (80\%)\\
    \hline
    All & 72 & 18 (25\%) & 31 (43.1\%) & 39 (54.2\%) & 50 (69.4\%)\\
    \hline
    \multicolumn{6}{l}{\% shows the percentage of bugs localized within the Top-{\em n}}\\
\end{tabular}
\end{center}
\vspace{-2em}
\end{table*}

Table~\ref{tbl:ranking-result} displays the outcomes of localizing the ground 
truth functions of the bugs in Top-{\em n}. The results show that the bug
localization result using the test programs generated from {\it DPGen4JIT}
performs effectively on both systems studied. 

In the case of V8, the ground truth buggy functions are ranked at the top 
(i.e., Top-{\em 1}) in 4 out of 37 bugs (10.8\%), and for 9 out of 37 bugs (24.3\%), 
the ground truth is ranked in the top 5. In the case of SpiderMonkey, the ground truth 
buggy functions are ranked at the top (i.e., Top-{\em 1}) in 14 out of 35 bugs (40\%), 
and for 22 out of 35 bugs (62.9\%), the ground truth is ranked in the top 5.

Overall, 25\%, 43.1\%, 54.2\%, and 69.4\% of ground truth functions 
are ranked within Top-{\em 1}, Top-{\em 5}, Top-{\em 10}, and Top-{\em 20}, 
respectively, using the test programs produced with our 
approach. Particularly, more than 44.44\% of the bugs can be localized 
within the Top-{\em 5}, which is the most preferred ranking that developers
expect from the bug localization approach to isolate the bug if they 
are unable to localize it to Top-{\em 1}.

\begin{figure}
  \centering
  \includegraphics[width=\linewidth]{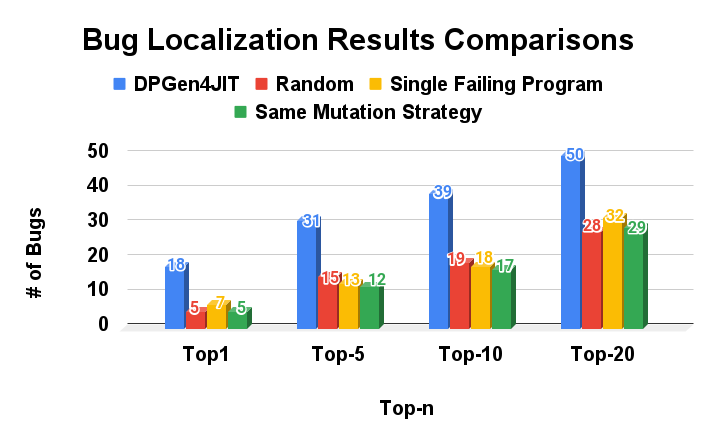}
  \caption{Bug localization result comparisons}
  \label{fig:compare}
\end{figure}

To assess the efficacy of the test programs produced using our approach,
we assessed the same 72 bugs with the current state-of-the-art techniques
({\bf Research Question 3}), focusing on the following:
(1) random generation of test programs 
by mutating the seed program randomly, similar to \cite{LimD21};
(2) retaining a single failing input (the original seed program) and
only generating passing test programs,
similar to \cite{HSZHZ19,ChenMZ20,DBLP:journals/tr/ZhouJRCQ22}; and
(3) generating both failing and passing test programs in the same way,
similar to \cite{DBLP:conf/issta/RobetalerFZO12}.

\subsubsection{Directed Generation vs. Random Generation}\label{sec:vsrandom}

Using the random test program generation, we located the bugs in 
the same manner as our approach using the Ochiai formula.
We performed 3 runs for each bug and then selected the median outcome, 
consistent with how we selected the results obtained from our tool.
Figure~\ref{fig:compare} shows the bug localization result comparison 
between the procedure performed with the test programs from {\it DPGen4JIT}
and the random program generator~\cite{LimD21}. The result shows that using
the test programs generated from our tool significantly outperforms
the result using the randomly generated programs. 
Our results demonstrate that bug localization using test programs generated by our approach outperformed the use of randomly generated test programs. Specifically, our approach was able to localize 18 bugs to Top-{\em 1} and 31 bugs to Top-{\em 5}, while the randomly generated test programs localized only 5 bugs to Top-{\em 1} and 15 bugs to Top-{\em 5}. Similar trends were observed for Top-{\em 10} and Top-{\em 20}, with our approach showing higher performance. 

The primary factors contributing to the outperforming bug localization 
results achieved by employing the test programs from {\it DPGen4JIT} 
compared to those generated randomly are twofold. Firstly, the randomly 
generated test programs lack sufficient diversity. We observed that 
the random generator produced several new programs by mutating 
the same code fragment with different values. While these new programs 
are not duplicates since the mutated values are distinct, the mutated
code segment is not relevant to the bug in the JIT compiler, rendering 
them less useful for bug localization. Secondly, the generator fails to 
produce passing programs for certain bugs, i.e., all test programs used
in the bug localization are failing programs. As a result, no comparison
points are available to localize the bugs accurately.

\subsubsection{Single vs. Multiple Failing Programs}\label{sec:1vsmany}

Several studies on generating test programs for 
bug localization focus on generating passing programs while having 
a single failing program, i.e., the seed program. 
\cite{HSZHZ19,ChenMZ20,DBLP:journals/tr/ZhouJRCQ22}
However, we propose that additional failing programs, generated with
guidance, can further increase the precision of bug localization.
To assess the effectiveness of our approach, we conducted an experiment 
involving a single failing seed program and a set of {\it N} passing programs. 
These passing programs were derived from the seed program by mutating 
the values of the AST nodes while maintaining the structure consistent. 
Furthermore, we ensured that the generated programs were distinct, 
meaning there were no duplicates among the new set of {\it N} programs.

Figure~\ref{fig:compare} shows the result of the ``Single Failing Program'' 
experiment compared to others. The results show that incorporating multiple
well-generated failing programs alongside the seed failing program during 
bug localization can substantially enhance accuracy. 
Specifically, the single failing program approach was able to localize 
7 bugs to Top-{\em 1} and 13 bugs to Top-{\em 5}, which is 
significantly lower than our approach. This trend was consistent 
across Top-{\em 10} and Top-{\em 20}, with our approach consistently 
outperforming.

The reason the single failing program approach performs worse
is its inability to
eliminate non-suspicious functions from the seed execution as discussed
in Section~\ref{sec:motivation}.

\subsubsection{Similarity of Passing/Failing Programs to the Seed}\label{sec:samevsdistinct}
While some approaches suggest generating both failing and passing programs
in the same manner, such as identifying bug-related parts of the seed program
and mutating them~\cite{DBLP:conf/issta/RobetalerFZO12}, 
our method differs by generating passing and failing test 
programs in different ways. Specifically, we aim to make passing 
programs as similar as possible to the seed program while making failing programs as 
dissimilar as possible.
To measure the performance of our idea to this opposing idea, we conducted 
an experiment with our altered tool to generate test programs by only
mutating the identified AST nodes for both passing and failing programs.

Figure~\ref{fig:compare} shows the result of 
the ``Same Mutation Strategy''
experiment compared to others. The results show that using a distinct
approach to generate different types of test programs resulted in 
a higher bug localization performance.
significantly improved indicate that our approach is more
effective than the same way mutation approach in localizing 
JIT compiler bugs.
The same mutation strategy was able to localize 
5 bugs to Top-{\em 1} and 12 bugs to Top-{\em 5}, which is 
significantly lower than our approach. This trend was consistent 
across Top-{\em 10} and Top-{\em 20}, with our approach consistently 
outperforming.

According to our experiment, there were two main factors that contributed
to the shortfall. Firstly, when we mutated the code fragments related to 
a bug to generate additional failing programs, we sometimes ended up with
passing programs instead, which led to generating only a minimal number
of failing programs. Secondly, due to the limited variety in the failing 
programs, the bug localization analysis failed to assign appropriate 
suspicious values to the executed instructions accurately. To address 
the aforementioned shortcomings, one possible solution is to generate 
a large number of test programs. For instance, statistical debugging
methods typically produce hundreds or even thousands of test programs.
Nonetheless, this approach often suffers from efficiency issues.

\subsection{Efficiency of {\it DPGen4JIT}}\label{sec:efficiency}

We evaluated the efficiency of {\it DPGen4JIT} by measuring the time 
it takes to generate a new set of test programs from a single seed 
test program in minutes. We used a user-specified value of $N = 30$.
The value 30 was selected arbitrarily and is further discussed 
in Section~\ref{sec:number-of-programs}.
On average, our tool took 2 minutes and 51 seconds to generate 
test programs, while the longest and shortest times were 4 minutes 
and 28 seconds and 1 minute and 42 seconds, respectively.

\section{Discussion}\label{sec:discussion}

\subsection{Enhanced Target Identification}\label{sec:enhanced-rules}

Our target identification method is able to successfully identify 
input code fragments related to a bug in the JIT compiler. 
However, we believe there is still room for improvement. 
Currently, the target identification process identifies code 
fragments related to a bug in isolation, even though the bug may 
be caused by a combination of multiple factors. 
For example, the JavaScript code snippet `{\tt let y = x + -0}'
triggers a bug in the JIT compiler optimization\cite{913296v8,1199345v8}.
The bug occurs only when the unary subtract operator is 
used with zero (i.e., `{\tt -0}'). The bug will not be triggered if 
the code is modified by replacing `{\tt -}' with another unary operator 
or zero with another number. While our tool can identify `{\tt -}' and `{\tt 0}' 
as the targets, it is not able to recognize that the bug is actually caused by 
the combination of these two code fragments. We believe that by enabling 
the tool to analyze input code and establish dependencies between 
the identified targets, considering the information about their combination, 
we can produce more sophisticated test programs that further increase 
the accuracy of bug localization.


\subsection{Number of Test Programs}\label{sec:number-of-programs}

It is possible that adding more failing and passing test programs 
beyond a certain number does not provide additional information that 
can further distinguish between the suspicious and non-suspicious
functions. This is because the added test programs may not reveal 
new behavior of the program, but rather confirm what was already 
discovered by the previous set of test programs.
This is because there could be redundant or similar test cases in 
the added test programs, which may not contribute to revealing new information 
about the program. Therefore, our future work is to find a way 
to automatically make the tool decide when to stop generating more failing 
test programs and passing programs.

\subsection{Generalizability of the Approach}\label{sec:generalizability}
While our approach was specifically designed for JIT compilers, 
we believe it can be extended to other software applications.
The key requirements for inputs are: syntactic structure that can be specified
using a context-free grammar (and therefore a parser
to read inputs into ASTs and a writer to write ASTs out to syntactically
correct inputs); and semantic constraints, such as type rules, on legal programs.
The key requirement for the program to be debugged is a test oracle that
can be used to distinguish buggy executions from non-buggy ones.  
We are currently investigating ways to expand our approach to other 
applications beyond JIT compilers.

\subsection{Threats to Validity}\label{sec:threats}

\subsubsection{Internal Threats}\label{sec:internal}

To ensure ranking accuracy, we conducted a comparison 
between our results and the ground truths. For real-world bugs, 
we carefully analyzed the bug reports, paying special attention to 
the ground truths, which we identified by studying the code changes 
and the developers' associated comments and discussions. When introducing 
bugs, we focused on the functions affected by the bug. 
Furthermore, we introduced the bugs by referring to bug reports to 
ensure that they had similar or identical characteristics to 
the reported bugs. Nevertheless, we intend to expand our experiments 
to include more real-world dynamic code generation bugs.

Another internal threat is the selection bias of existing approaches 
used for comparison. To address this, we reviewed related works with similar 
objectives, which involved generating test programs for bug localization. 
We selected three different approaches to evaluate the same bugs as in 
our experiment. Additionally, we plan to test our approach on different 
bugs and explore different comparison methods to reduce the impact of 
selection bias.

\subsubsection{External Threats}\label{sec:external}

To strengthen the external validity of our study, we acknowledge 
the potential difficulty in applying our approach to various JIT compilers, 
which is a primary risk from an external standpoint. To mitigate this risk, 
we conducted thorough testing of our findings using bugs found in 
two popular JIT compilers: Google's V8/Turbofan and Mozilla's 
SpiderMonkey/IonMonkey. However, we recognize that there may be 
variations in JIT compilers that could affect the applicability of 
our approach. To further address this risk, we plan to conduct further 
experiments with a wider range of bugs on other JIT compilers, 
such as Apple's JavaScriptCore/DFGJIT, to ensure that our approach 
can be generalized to different JIT compilers.

\section{Related Work}\label{sec:related-work}
While the test program generation proposed by Lim and Debray~\cite{LimD21} 
shares the goal of generating test programs for localizing bugs in 
the JIT compilers, it has several limitations. In their approach, 
AST nodes are randomly selected and mutated for a user-specified 
number of times. This does not guarantee the generation of quality 
test programs that can effectively eliminate unnecessary program entities.

The approaches~\cite{HSZHZ19,ChenMZ20,DBLP:journals/tr/ZhouJRCQ22} 
for generating test programs for traditional compilers, such as GCC 
and LLVM, typically focus on generating only the passing test programs. 
As discussed in Section~\ref{sec:evaluation}, it is more advantageous 
to create additional failing test programs to eliminate less suspicious 
program entities, which could potentially lead to a more accurate 
bug localization. A recent study~\cite{chen2023compiler} that employs 
machine learning techniques to generate test programs is designed 
to detect new bugs, similar to fuzzers.

Blazytko {\em et al.} employed AFL fuzzer~\cite{afl} with 
a specialized configuration (i.e., {\it crash exploration} mode) 
to produce test programs. Nevertheless, utilizing fuzzers
\cite{afl,wangfuzzjit,ruderman2007introducing,honggfuzz}, 
whose primary goal is to identify new bugs by examining the execution 
coverage, may not be the best option for bug localization. This is because 
the newly generated test programs may vastly differ from the seed and not 
link to the bug of interest we want to locate. Additionally, solely 
considering buggy behavior as a crash can result in the misclassification 
of test programs. For instance, if a program produces incorrect output
but terminates without crashing, it may be mislabeled as a passing program 
because it did not crash.

The approaches~\cite{DBLP:conf/issta/ArtziDTP10,%
DBLP:conf/icse/BaudryFT06,DBLP:conf/kbse/CamposAFd13,%
DBLP:conf/wcre/LiuLNB17,DBLP:conf/issta/RobetalerFZO12} 
used for ordinary programs limits scaling to JIT compilers. 
This is because the input to the JIT compiler is another program, 
while these approaches target generating test cases, such as input 
values, or directly mutating the target system.

Mutation-based fault localization (MBFL)
\cite{DBLP:conf/kbse/HongLKJKKK15,DBLP:conf/icse/BaudryFT06,%
DBLP:journals/pacmpl/LiZ17,DBLP:conf/icst/MoonKKY14,%
DBLP:conf/icst/PapadakisT12,DBLP:journals/stvr/PapadakisT15,%
DBLP:conf/oopsla/Zhang0K13} 
involve mutating the program containing the bug in order to identify 
the program entities that are likely to be responsible. 
However, these approaches are challenging to scale for 
JIT compilers. Especially, JIT compilers are part of larger systems, 
e.g., JavaScript engines or Virtual Machines, that the approach 
attempting to mutate specifically targeting the JIT compiler alone 
is a non-trial task.

\section{Conclusion}\label{sec:conclusion}
Bug localization for JIT compilers relies on 
analyzing the execution behaviors of test programs.
However, current automatic test program generation approaches 
are not effective for JIT compiler bug localization. 
To address this, we developed a novel approach that generates 
effective test programs for JIT compiler bug localization.
We evaluated through experiments on widely-used 
JIT compilers, demonstrating its effectiveness.

\section*{Acknowledgements}\label{sec:ack}
This research was supported in part by the National Science Foundation 
under grant no. 1908313.

\bibliographystyle{IEEEtran}
\bibliography{references}

\end{document}